\documentclass[iop]{emulateapj}

\usepackage{apjfonts}
\usepackage{pslatex}

\usepackage{ulem}
\usepackage{color}
\usepackage{soul}
\usepackage{amsmath}
\usepackage{natbib,url}
\usepackage{textcomp}
\usepackage{amssymb}
\usepackage{hyperref}
\hypersetup{colorlinks=true, urlcolor=blue, citecolor=blue}

\shorttitle{Compact Jets and X-ray Variability Properties}
\shortauthors{Din\c{c}er et al.}

\begin{document}

\title{Complete Multiwavelength Evolution of Galactic Black Hole Transients During Outburst Decay II: Compact Jets and X-ray Variability Properties}
\author{T. Din\c{c}er\altaffilmark{1}, E. Kalemci\altaffilmark{1}, J. A. Tomsick\altaffilmark{2}, M. M. Buxton\altaffilmark{3}, C. D. Bailyn\altaffilmark{3}}
\affil{}
\altaffiltext{1}{Faculty of Engineering and Natural Sciences, Sabanc{\i} University, Orhanl{\i}-Tuzla 34956, \.{I}stanbul, Turkey}
\altaffiltext{2}{Space Sciences Laboratory, 7 Gauss Way, University of California, Berkeley, CA 94720-7450, USA}
\altaffiltext{3}{Astronomy Department, Yale University, P.O. Box 208101, New Haven, CT 06520-8101, USA}

\begin{abstract}
We investigated the relation between compact jet emission and X-ray variability properties of all black hole transients with multiwavelength coverage during their outburst decays. We studied the evolution of all power spectral components (including low frequency quasi-periodic oscillations), and related this evolution to changes in jet properties tracked by radio and infrared observations. We grouped sources according to their tracks in radio/X-ray luminosity relation, and show that the standards show stronger broadband X-ray variability than outliers at a given X-ray luminosity when the compact jet turned on. This trend is consistent with the internal shock model and can be important for the understanding of the presence of tracks in the radio/X-ray luminosity relation. We also observed that the total and the QPO rms amplitudes increase together during the earlier part of the outburst decay, but after the compact jet turns either the QPO disappears or its rms amplitude decreases significantly while the total rms amplitudes remain high. We discuss these results with a scenario including a variable corona and a non-variable disk with a mechanism for the QPO separate from the mechanism that create broad components. Finally, we evaluated the timing predictions of the magnetically dominated accretion flow model which can explain the presence of tracks in the radio/X-ray luminosity relation.
\end{abstract}

\keywords{accretion, accretion disks -- black hole physics -- stars: winds, outflows -- X-rays: binaries}

\section{Introduction}
Jets constitute an integral part of the accretion flow in black hole systems of all sizes, and there are many open questions about how they form and affect their environment. Since the jet emission is prominently detected in radio and infrared, and the accretion flow is often characterised by X-ray spectral and timing properties, a systematic multiwavelength approach is required to understand accretion/ejection relation quantitatively. Our group has been working on this problem by characterizing the X-ray properties while the jets turn on (determined through changes in radio and/or infrared emission) during outburst decays of Galactic black hole transients (GBHT). In \cite{Kalemci13}, which will be referred as Paper I from hereon, we concentrated on the changes in X-ray spectral properties, and determined the timescales of important transitions during the outburst decay before, during and after the formation of the compact jets. This is the second part of the systematic study for which we concentrate on the changes in short term X-ray variability properties during the same time. Earlier observational studies that tie jets to timing properties for individual sources exist \citep{Miller12,Yan13}, and there is one systematic study that relates behavior of short term variability properties to optically thin radio flares \citep{Fender09}. To the best of our knowledge, there is no detailed systematic study with a large sample of outburst decays that are analyzed uniformly that investigates the relation between the compact jet formation and properties of both the broad and narrow components in the power spectra of GBHTs as we attempt in this work.

GBHT are accreting binary systems which spend most of their time in quiescence and undergo occasional X-ray outbursts lasting for months to years \citep{Tanaka96}. During these outbursts, the GBHTs exhibit two main states identified by their X-ray spectral and temporal properties. In the soft state the X-ray spectrum shows strong thermal emission from an optically thick, geometrically thin disk \citep{Shakura73} and the variability is low ($\sim$a few \% rms amplitude) whereas in the hard state the X-ray spectrum is much harder, dominated by emission in the form of a power-law from a hot inner flow (ADAF or corona) and/or the base of a jet at luminosities $<$ $10^{-3} L_{Edd}$ \citep{Russell10}, the variability is much stronger ($>$20 \% rms amplitude), and low frequency quasi-periodic oscillations (QPO) with frequencies ranging from a few mHz to $\sim$10 Hz are commonly observed \citep{Remillard06}. There are also intermediate states occurring between the soft and the hard states for which the X-ray spectral properties show the characteristics of a mixture of both states. See \cite{Belloni10} for a recent review of the states and \cite{Done07} for a detailed discussion of the accretion models.

These states are closely related to the behavior of compact, steady jets. They are commonly observed in the hard state as a persistent radio emission with a flat or slightly inverted spectrum, and also sometimes as an excess of infrared emission \citep{Corbel00, Fender01, Buxton12, Dincer12}. In the soft state, in contrast, they are thought to be no longer present because the emission in the radio and infrared bands are highly quenched \citep{Homan05, Russell_q11}.

A strong correlation exists between the radio and X-ray luminosity of GBHTs during the hard state. This correlation implies a connection between the jet and the accretion flow (optically thick disk or hot inner flow). The correlation has two tracks: ``standards'' and ``outliers''. While both tracks exhibit a power-law relation their slopes are different, the outliers show fainter radio luminosities for the same X-ray luminosities than the standards \citep{Gallo12, Corbel13}. At $L_{3-100 keV}$ $<$ $5\times10^{-3}L_{Edd}$, some sources in the outliers track approach the standard track in the radio/X-ray luminosity plane \citep{Coriat11, Ratti12}.
To date, several studies have suggested physical explanations for the presence of the two different tracks in the radio/X-ray luminosity relation. These explanations include differences in the strength of the jet magnetic field \citep{Casella09}, the radiative efficiency of the accretion flow \citep{Coriat11}, presence/absence of a cool inner disk \citep{Meyer-Hofmeister14}, or a magnetically dominated accretion flow \citep[MDAF,][]{Meier12}. \cite{Soleri11} tried relating the tracks to binary parameters and presence or lack there of transitions to the soft state, but failed to find a dependence. 

In this work, we present a systematic multiwavelength analysis of the GBHTs focusing on the relation between the jet emission and X-ray power spectral properties. The sample is the same as the one used in Paper I which includes the outburst decays of 4U 1543-47 in 2002, GRO J1655-40 in 2005, GX 339-4 in 2003, 2005, 2007, 2011, H1743-322 in 2003, 2008, 2009, XTE J1550-564 in 2000, XTE J1720-318 in 2003, and XTE J1752-223 in 2010.

\section{Data Analysis}
\label{sec:dataan}
We used archival Rossi X-ray Timing Explorer ({\it{RXTE}}, \citealt{Bradt93}) PCA data for all timing analysis, while the X-ray spectral analysis made use of both the PCA and HEXTE instruments in Paper I.

\subsection{X-Ray Timing Analysis}
For each source, we obtained light curve with a time resolution of 2$^{-11}$ s in the 3--25 keV energy band. We divided the light curve from high resolution data into 16 s segments for bright observations, generated a power-density spectrum (PDS) from each segment, and averaged them together. As the source flux decreases during outburst decays, we progressively increased the segment length up to 128 s. We normalized the PDS as in \cite{Miyamoto89} and corrected the dead time effects using the model of \cite{Zhang95} with a dead time of 10 $\mu$s per event.

We fitted the power spectra by a number of Lorentzian profiles, each described by their characteristic frequency $\nu_{max}=(f_0^2+\Delta^2)^{1/2}$, where $f_0$ is the centroid frequency, $\Delta$ the HWHM, and by the quality factor ($Q\equiv f/2\Delta$). Lorentzians with $Q$ $>$ 2 were referred as QPOs and those with $Q$ $<$ 2 as broad noise components. At low flux levels, we multiplied the total fractional rms by a correction factor to eliminate the reduction by the Galactic ridge emission (see Paper I).

When necessary, we estimated upper limits on the QPO rms amplitudes using the method given in \cite{Vaughan11}. In this method, we compared two hypothesis, the null hypothesis $H_0$ of a continuum spectrum described by a number of broad Lorentzians, and the alternative hypothesis $H_1$ which includes an additional narrow Lorentzian with a fixed quality factor of $Q=10$ for the QPO. The upper limits are given for the chance of a false detection probability of $\alpha=0.01$, and the probability of a detection of $\beta=0.5$ (see \citealt{Vaughan11} for definition of $\alpha$ and $\beta$ parameters).

\subsection{X-Ray Spectral Analysis}
We adopted the X-ray spectral parameters presented in Paper I. These parameters were obtained by fitting the X-ray spectrum with a phenomenological model consisting of photoabsorption, a smeared edge, a multicolor disk blackbody, a power-law, a narrow Gaussian to model the iron line, and, if necessary, an additional power-law representing the Galactic ridge emission. The Eddington Luminosity Fraction (ELF) for each observation was calculated as the ratio of 3--200 keV luminosity to the Eddington luminosity as in Paper I . 

\section{Results}
We present our results based on the three transitions identified in Paper I, the timing transition (TT), the index transition (IT), and the compact jet transition (CJT).  The TT is defined as a sudden increase in the total rms amplitude accompanied by an increase in the power-law flux \citep{Kalemci04}, the IT is defined as the start of the hardening of the X-ray spectrum, and the CJT is defined as the start of the increase in the NIR flux or the time of the first optically thick radio detection. For details of how these transition times are identified, see Paper I.

In figures, we use the color scheme used in Paper I to show the observations before and after the transitions in the color online version. The observations between the TT and the IT are shown with green, the observations between the IT and the CJT with blue, and the observations after the CJT are shown with red.

\begin{figure}
 \epsscale{1.0}
 \plotone{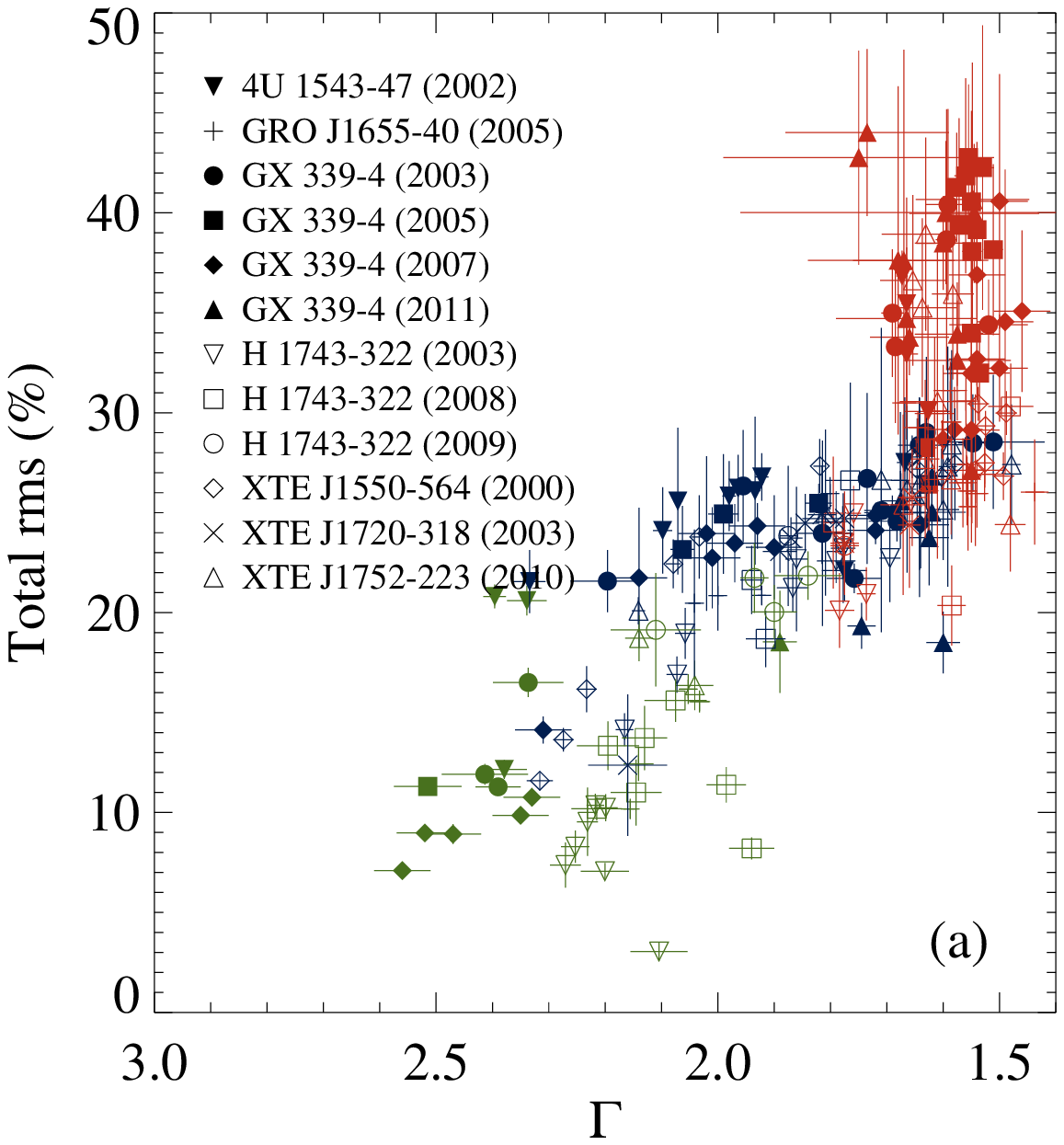}
 \plotone{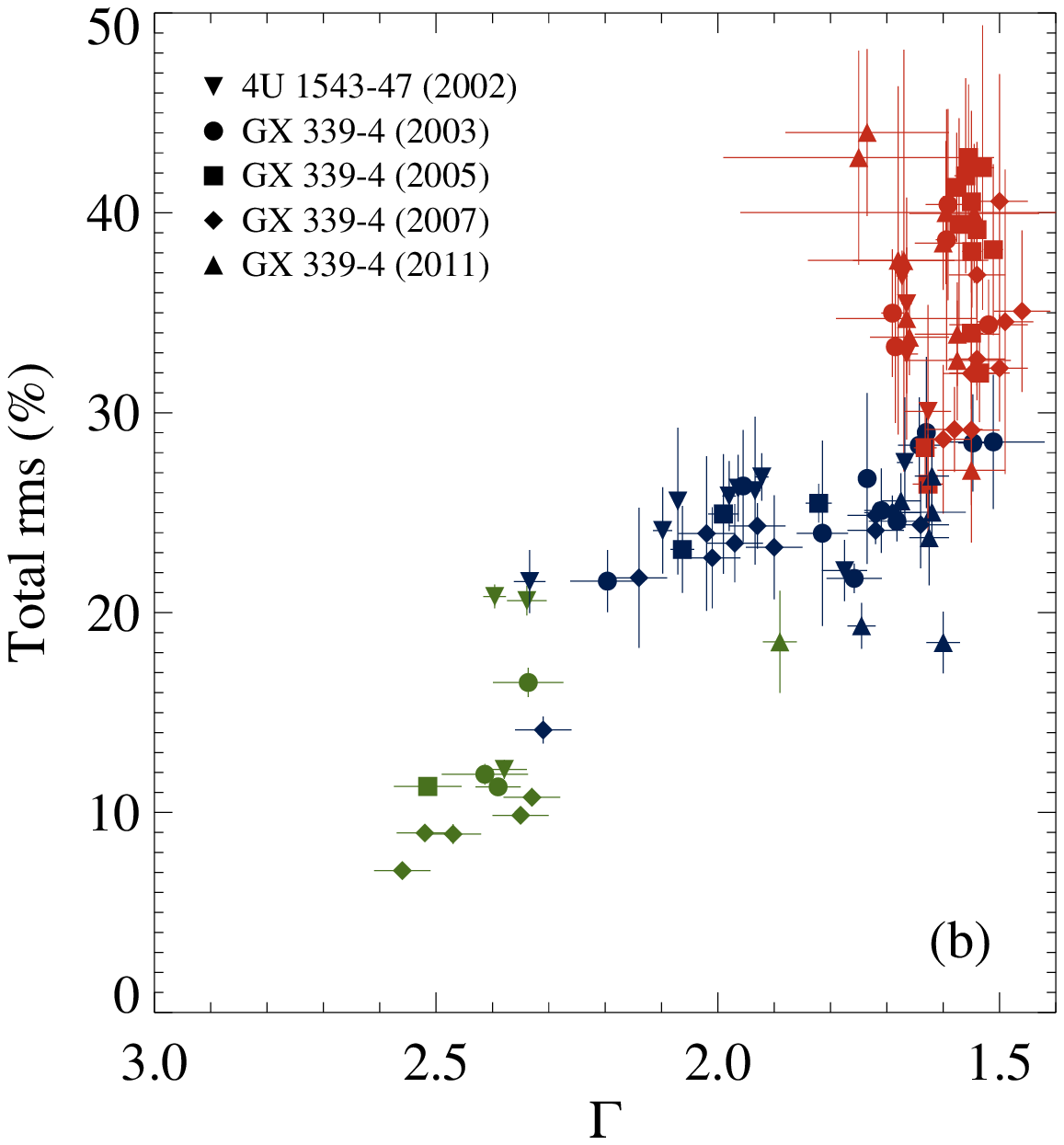}
 \plotone{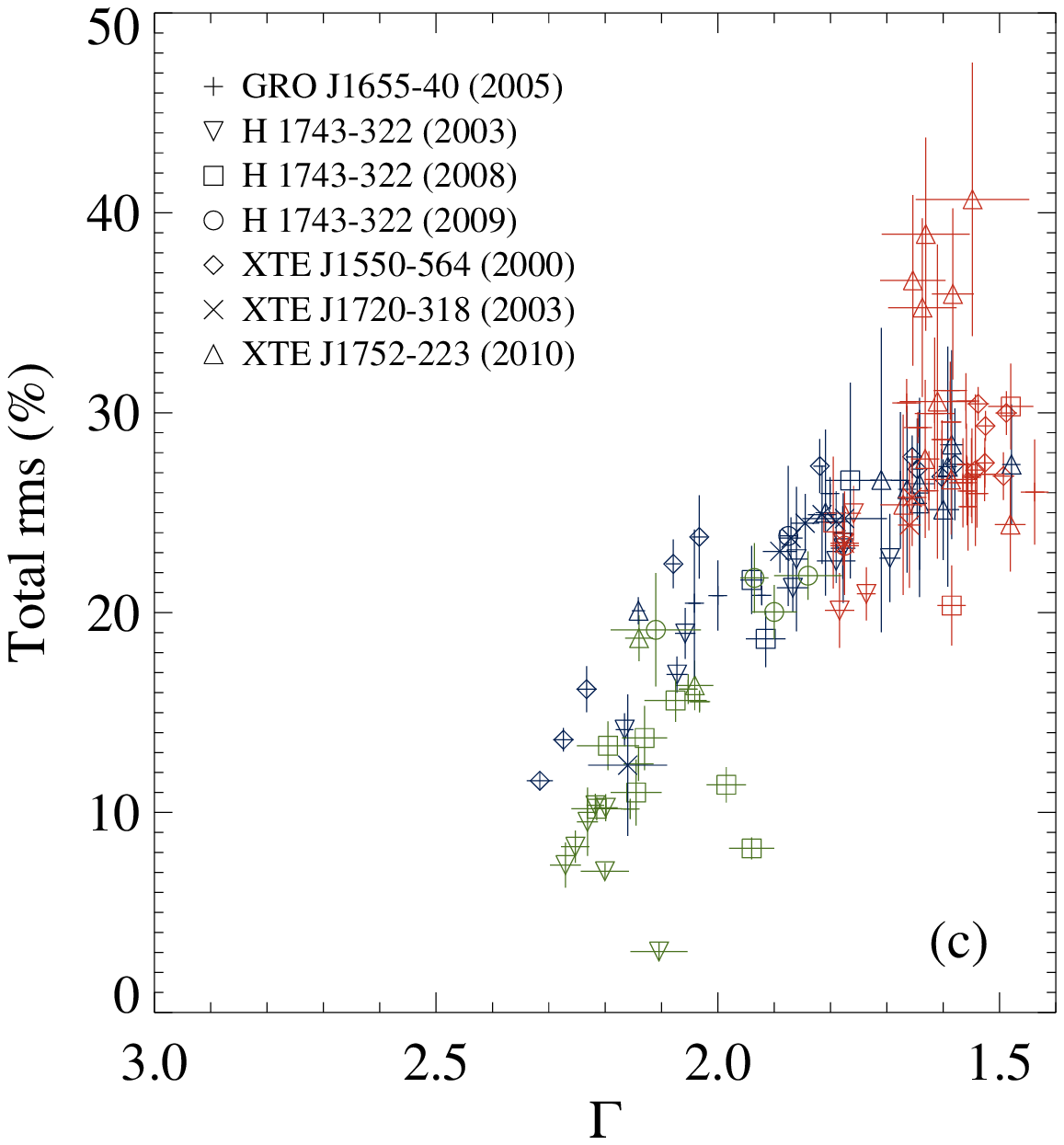}
 \caption{Rms amplitude of variability as a function of photon index ($\Gamma$) for the observations of a) all the sources, b) the standards, and c) the outliers. In the color online version, the observations between the TT and the IT are shown with green, the observations between the IT and the CJT with blue, and the observations after the CJT are shown with red.}
 \label{fig:Gammarms}
\end{figure}

\begin{figure}
 \epsscale{1.0}
 \plotone{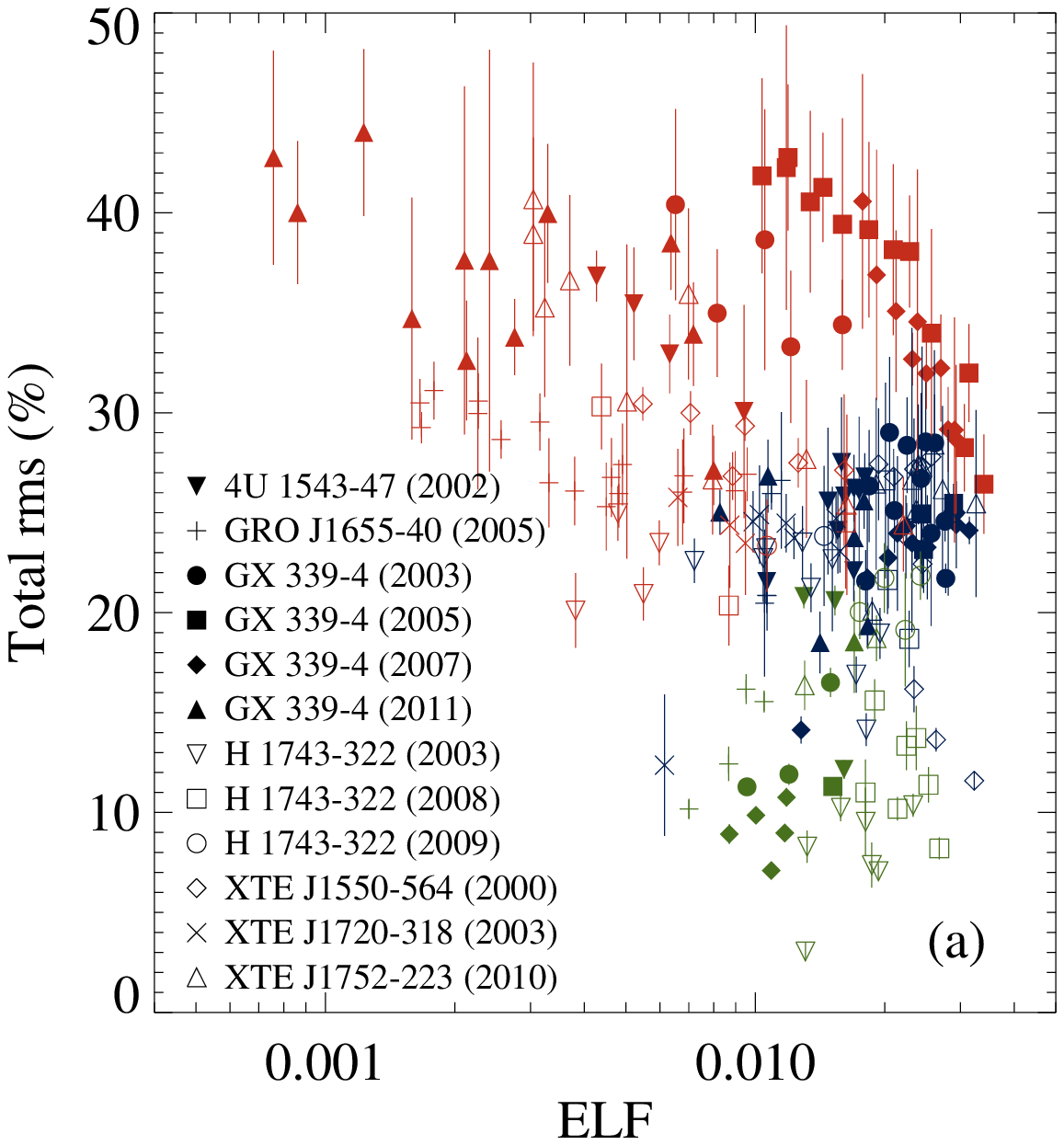}
 \plotone{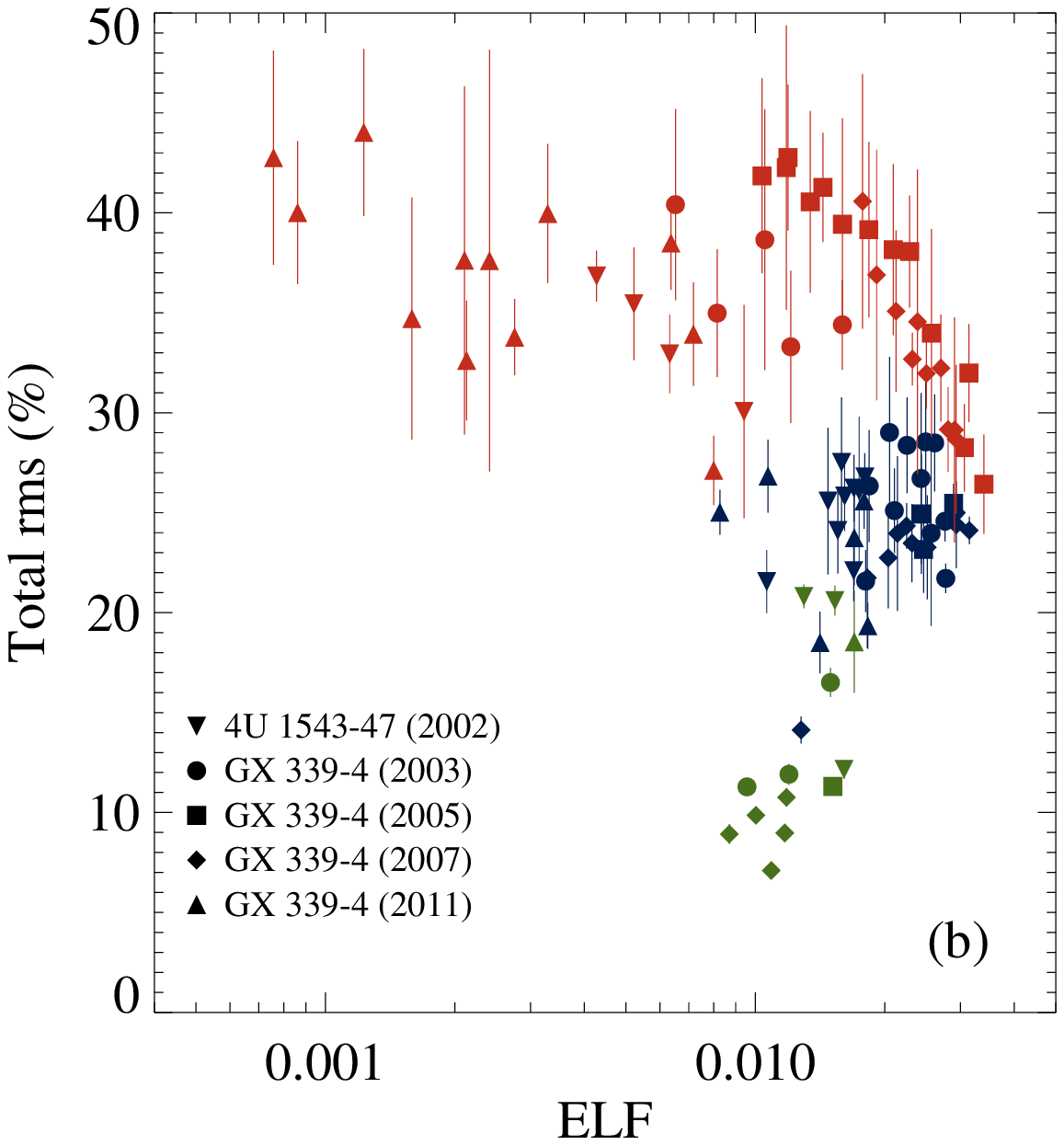}
 \plotone{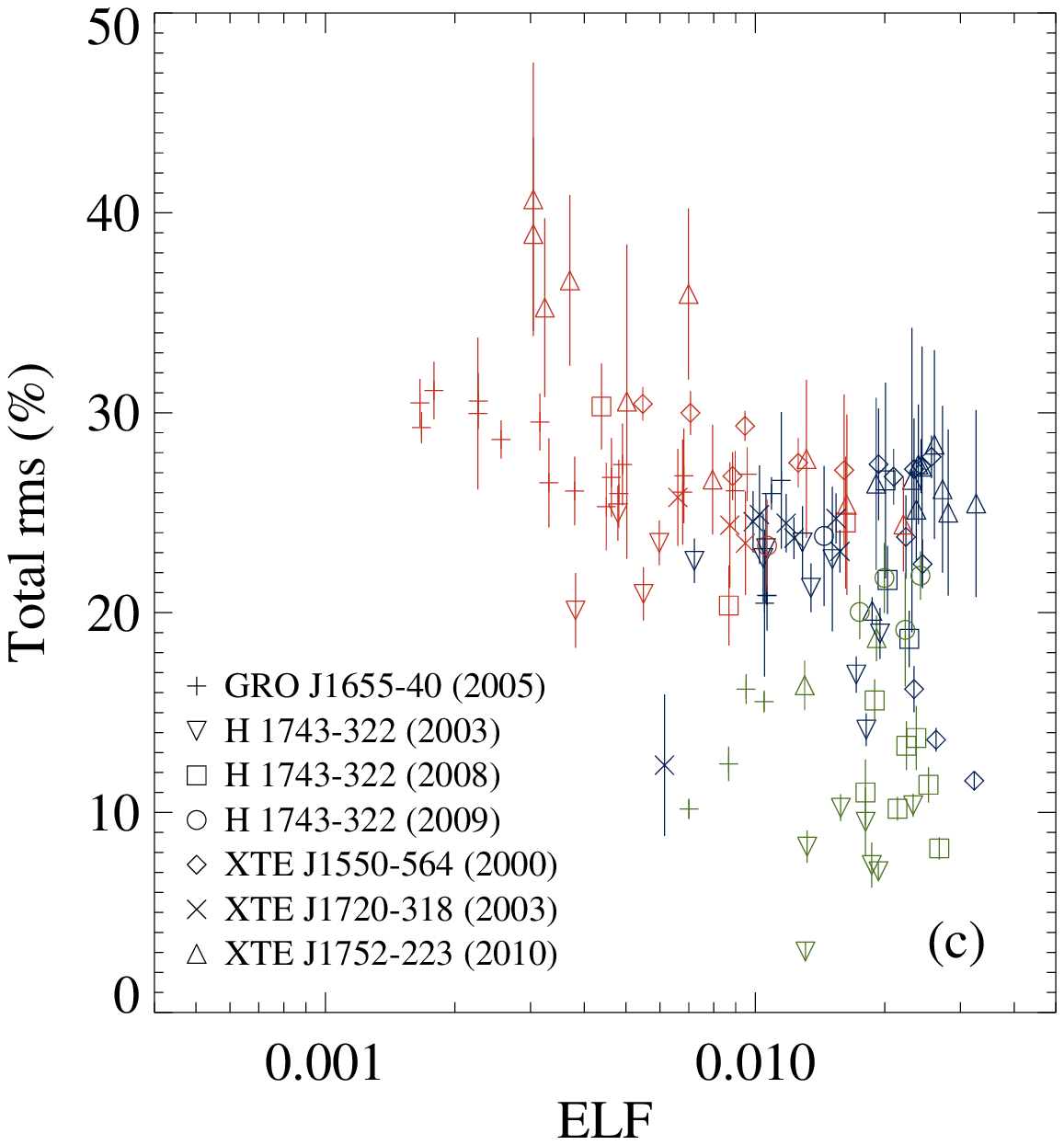}
 \caption{Rms amplitude of variability as a function of ELF for the observations of a) all the sources, b) the standards, and c) the outliers.}
 \label{fig:Lrms}
\end{figure}

\subsection{Rms Variability: The Standards vs The Outliers}
\label{sec:varoftrack}
\begin{figure}
 \epsscale{1.0}
 \plotone{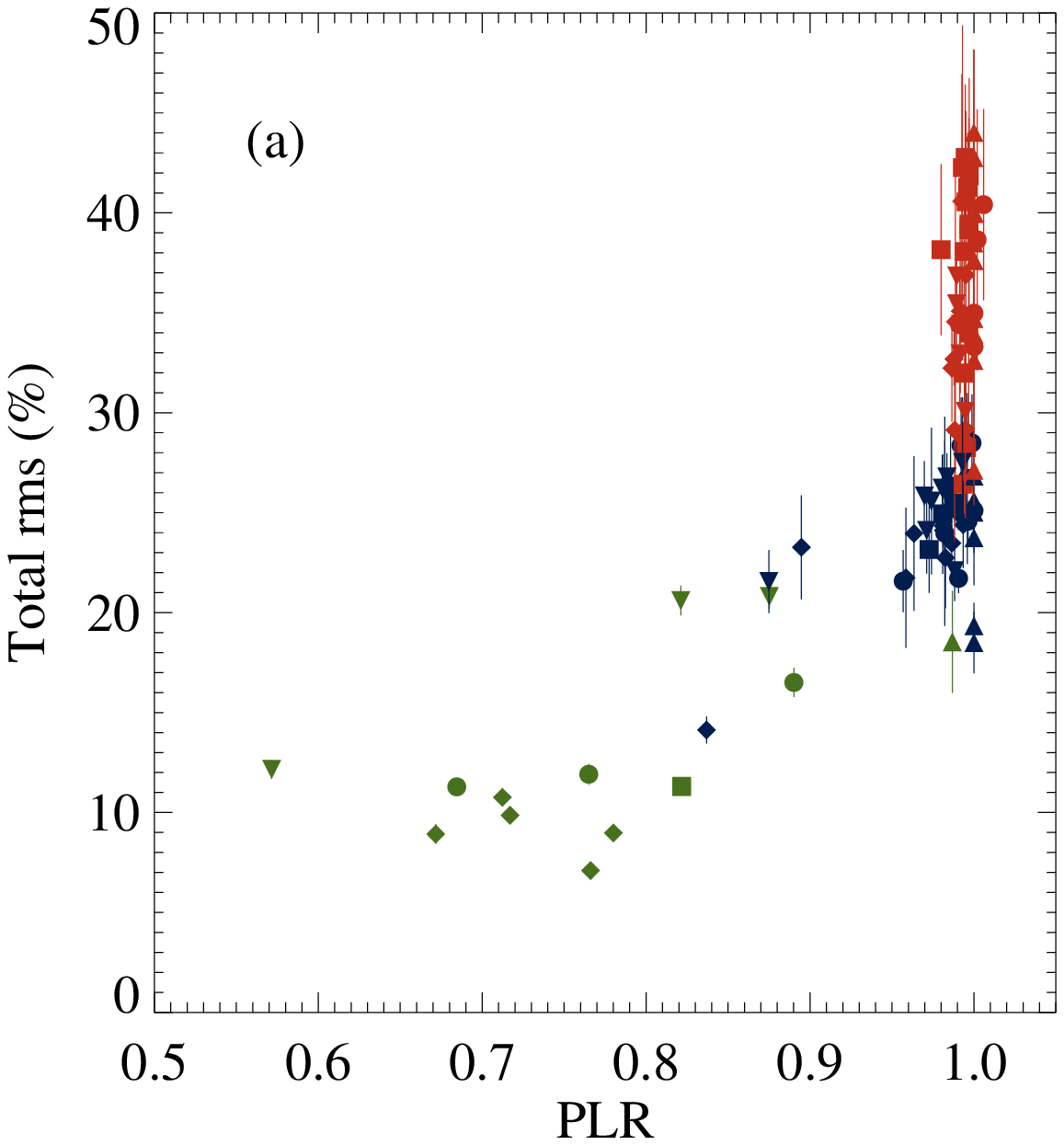}
 \plotone{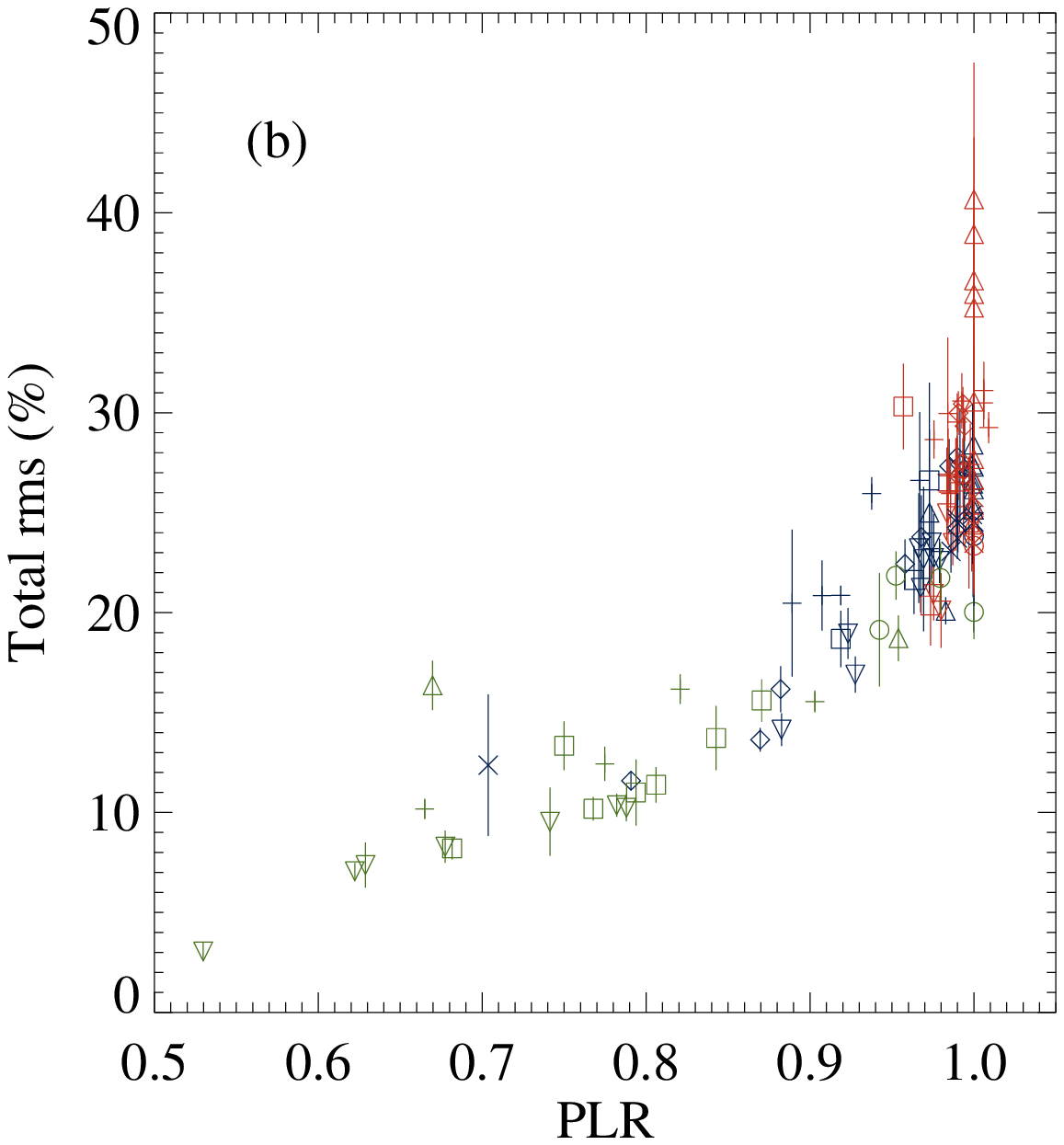}
 \caption{Total rms amplitude of variability as a function of PLR for a) the standards and b) the outliers.}
 \label{fig:rmsplr}
\end{figure}

In this section, we compare the X-ray timing properties of standards and the outliers of the radio/X-ray luminosity correlation during the formation and the evolution of the compact jets. In Figures~\ref{fig:Gammarms} and \ref{fig:Lrms}, we plot the total rms variabilities against the power-law indices and ELFs, respectively. Each figure consists of three panels. The panels from top to bottom show the data obtained from all the sources, the standards, and the outliers, respectively. In each panel, outburst decays are distinguished from each other with different plotting symbols. In Figures~\ref{fig:rmsplr}a and ~\ref{fig:rmsplr}b, we plot the total rms amplitude as a function the power-law ratio (PLR, the ratio of the power-law flux to the total flux in 3-25 keV energy band) for the standards and the outliers, respectively.

In Figure~\ref{fig:Gammarms}a, a global relation is observed for all GBHTs. As expected, at the beginning of the outburst decays, the spectral indices are soft, and the rms amplitudes are small. During the hardening the rms amplitudes gradually increase, and after the CJT the rms amplitudes take a sharp upwards turn, reaching above 40\% for some sources. In Fig.~\ref{fig:Lrms}a, we observe that at a few percent of the ELF, the rms amplitudes increase from a few to 25\%. As the ELF decreases further and the CJT occurs, we observe a large scatter in the rms amplitudes for a given ELF.
The same data, divided into two groups based on the association of the sources to the tracks observed in the radio/X-ray luminosity relation \citep[e.g. see the tracks with the sources in][]{Corbel13}, show differences between the behavior of the two groups. First, there are occasions that the CJT occurs along with softer spectral indices and lower rms amplitudes in outliers compared to standards (see Figures~\ref{fig:Gammarms}b,c). Second, the rms amplitudes after the CJT show two trends, the standards reach higher rms amplitudes with a weighted mean of 34.1 $\pm$ 0.4 \%, and the outliers stay constant at 23.1 $\pm$ 0.2 \% (see Figures~\ref{fig:Lrms}b and \ref{fig:Lrms}c).

XTE~J1752-223 has higher rms amplitude at low flux levels compared to the other outliers. We must note that low flux observations of this source are strongly affected by the Galactic ridge emission. We have incorporated the effect of ridge emission in the rms amplitude \citep{Chun13}, however the uncertainty in the ridge emission results in large errors in the rms amplitude. Therefore the high rms values from low luminosity XTE~J1752-223 should be regarded with caution.

We checked whether the highest rms amplitude of variabilities in XTE~J1752-223 corresponds to a transition from outlier track to standard track in the radio/X-ray luminosity relation. However, the observations with higher rms amplitude of variability (between MJD~55,333 and 55,353) belong to the times when the source is still in the outlier track \citep[see][]{Ratti12}. We note that we are not able to track X-ray timing information as the sources transition from the outlier track to the standard track. This is because of low source count rate at such low luminosities. 

Another difference in the rms amplitude of the standards and the outliers is seen in Figures~\ref{fig:rmsplr}a and \ref{fig:rmsplr}b. The rms amplitudes form a tight correlation with the PLR for the outliers, but are more scattered for the standards.

\begin{figure*}
\epsscale{1.1}
\plottwo{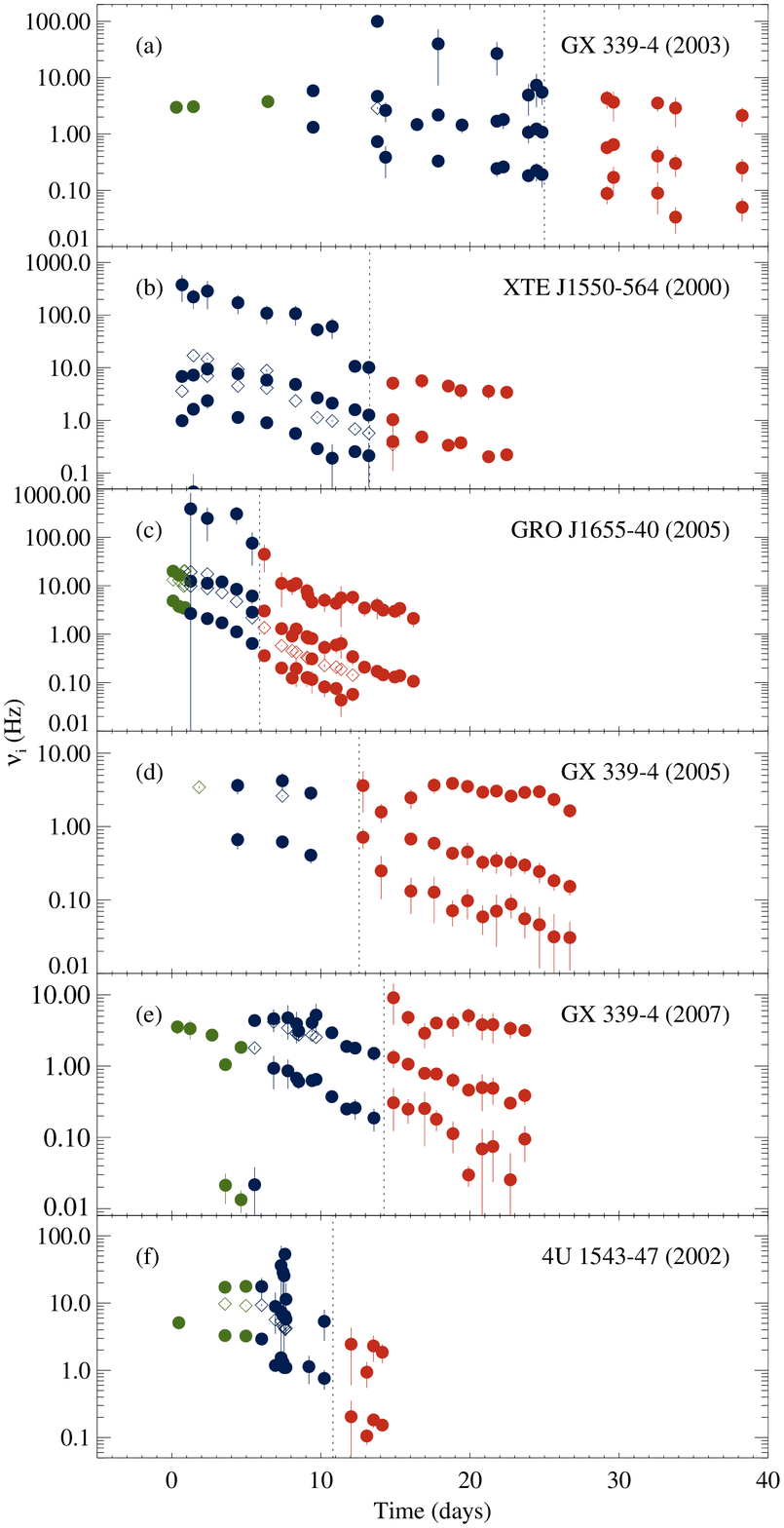}{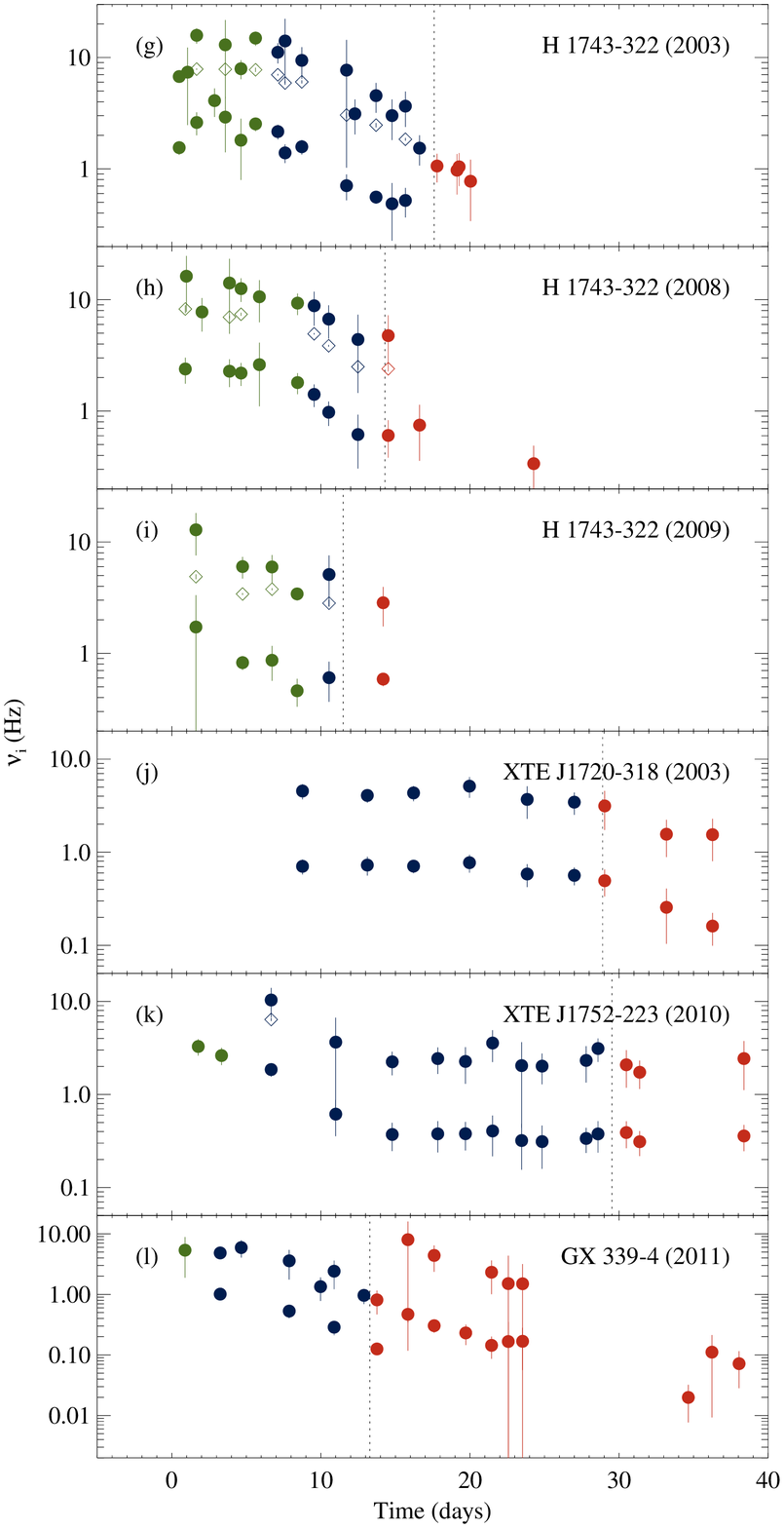}
\caption{Time evolution of the peak frequencies for all outburst decays. Time 0 denotes the time of the timing transition. Dotted lines show the time of the compact jet transition.}
\label{fig:timevolfreq}
\end{figure*}

\subsection{Time Evolution Of The X-ray Variability Properties}
Figure~\ref{fig:timevolfreq} shows the evolution of the peak frequencies for all outburst decays. The broad Lorentzian components identified in each observation are shown with solid circles, and QPO peak frequencies are shown with empty diamonds. The most noticeable result in the evolution of the peak frequencies is their decreasing trend in time which is a well known characteristic of the GBHTs during outburst decay \citep[e.g.][]{Kalemci03,Klein-Wolt08}. In the following, we characterise the behavior of the broad noise components and the QPO with respect to formation of compact jets (CJT) in detail.

\begin{figure}
\epsscale{1.2}
\plotone{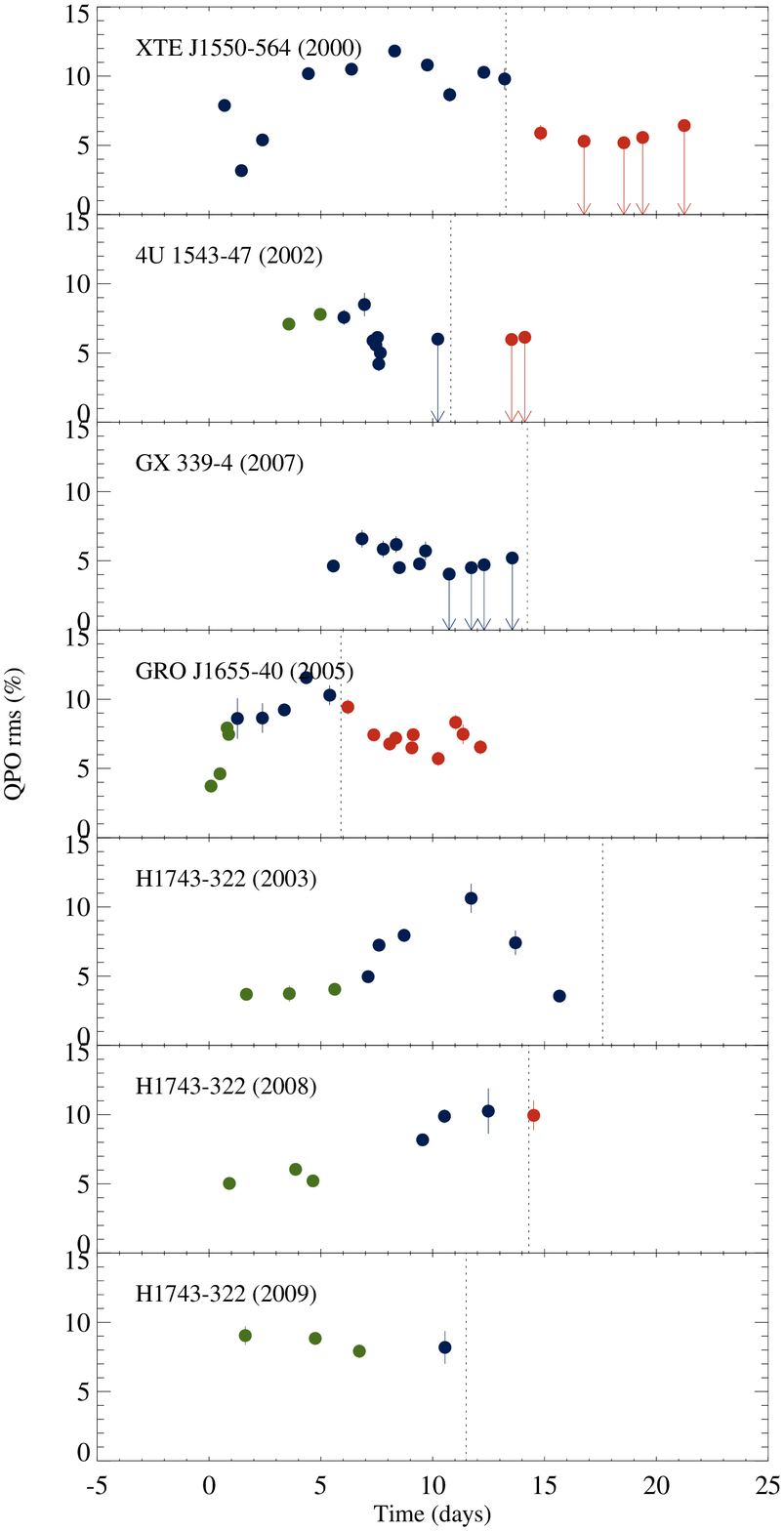}
\caption{Time evolution of the QPO rms amplitude of variability. Time 0 denotes the time of the timing transition. Dotted lines show the time beyond which the jet is optically thick.}
\label{fig:evolqporms}
\end{figure}

\begin{figure}
\epsscale{1.2}
\plotone{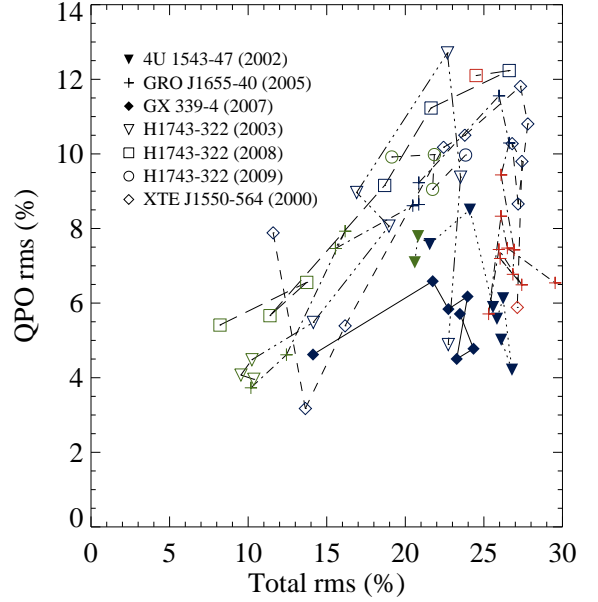}
\caption{Total vs QPO rms amplitude of variabilities.}
\label{fig:totqporms}
\end{figure}

\begin{figure}
\epsscale{1.2}
\plotone{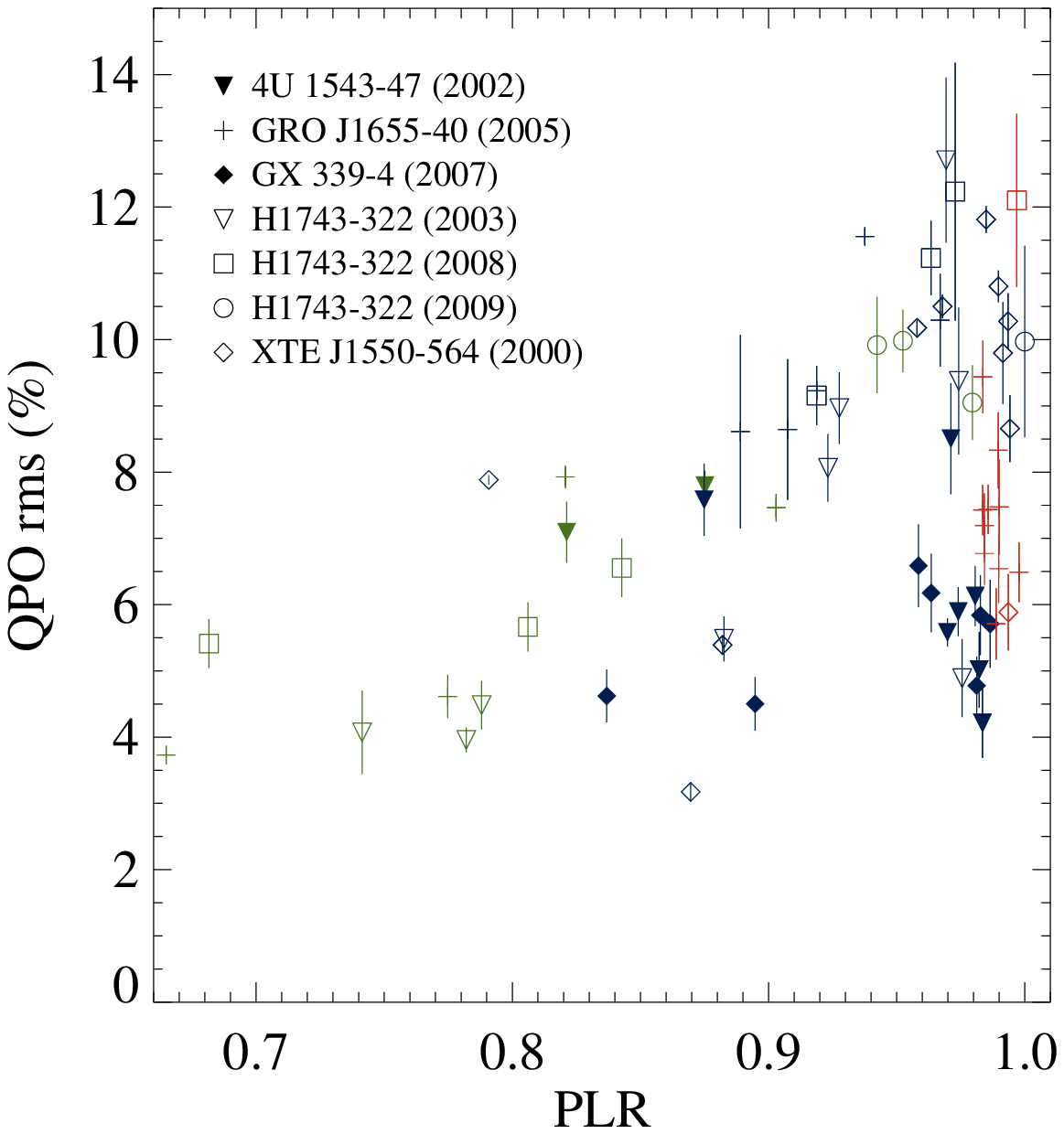}
\caption{QPO rms amplitude of variability as a function of PLR.}
\label{fig:qpormsplr}
\end{figure}

\subsubsection{Evolution of Broad Noise Components}
In most of the outburst decays, we do not observe sudden changes in the evolution of the peak frequencies. However when we observe changes, they do happen close to the CJT. For example, there is a sudden decrease in the peak frequencies of the highest frequency Lorentzian in GRO J1655-40 and XTE J1550-564 within 3 days prior to the CJT. The number of broad components increase for GX 339-4 (2005 and 2007) within 3 days after the CJT.

\subsubsection{Evolution of QPOs}
As shown in Figure~\ref{fig:timevolfreq}, we observe that in some outburst decays (4U 1543-47, GX 339-4 [2007], H1743-322 [2003, 2008, 2009], XTE J1550-564), QPOs are no longer required in the PDS fits after the CJT. This is more clearly observed in Figure~\ref{fig:evolqporms} where we plot the evolution of the QPO rms amplitudes as a function of time through different transitions. For XTE J1550-564, there is a sudden drop in the rms amplitude of the QPO at the CJT, and from there on it is no longer required in the fit, so we only have upper limits. For 4U 1543-47 and H1743-322 (2003), a sudden decrease in the rms amplitude occurs 4 days before the CJT, and the QPO is not required in the fits afterwards. For the 2008 and 2009 outburst decays of H1743-322, the QPOs are not required in the PDS fits after the CJT and their rms amplitudes did not show any decrease prior to the CJT. For the 2007 outburst decay of GX 339-4, the QPO is no longer required in the fits starting from 4 days prior the CJT. On the other hand, for GRO J1655-40, the QPO is observed throughout the outburst decay, its amplitude climaxing a few days before the CJT.

The evolution of the QPO rms amplitude is similar in H1743-322 (2003), GRO J1655-40, XTE J1550-564. The QPO rms amplitude first increases, then makes a peak and finally decreases. Other outburst decays exhibit only a portion of this trend. For instance, 4U 1543-47 shows only the peak and the decrease whereas H1743-322 (2008) shows only the increase and the peak. In Figure~\ref{fig:totqporms}, we plot the total rms amplitude against the QPO rms amplitude. At lower total rms amplitudes, a positive correlation exists between the two parameters. When the total rms amplitude reaches between 20-30\%, the QPO rms amplitude of variability decreases (or the QPO is not required in the fit). In Figure~\ref{fig:qpormsplr}, we plot the QPO rms amplitude against the PLR to see if this trend is related to the presence of disk emission. The increase in the QPO rms amplitude is associated with the increase in the PLR which is expected if the majority of the disk emission is not variable, but the decrease in the rms occurs at the highest PLR for which more than 95\% of the emission comes from the corona in the PCA band.

\subsection{$\nu_{qpo}$-Luminosity Relation}
\begin{figure}
 \epsscale{1.2}
 \plotone{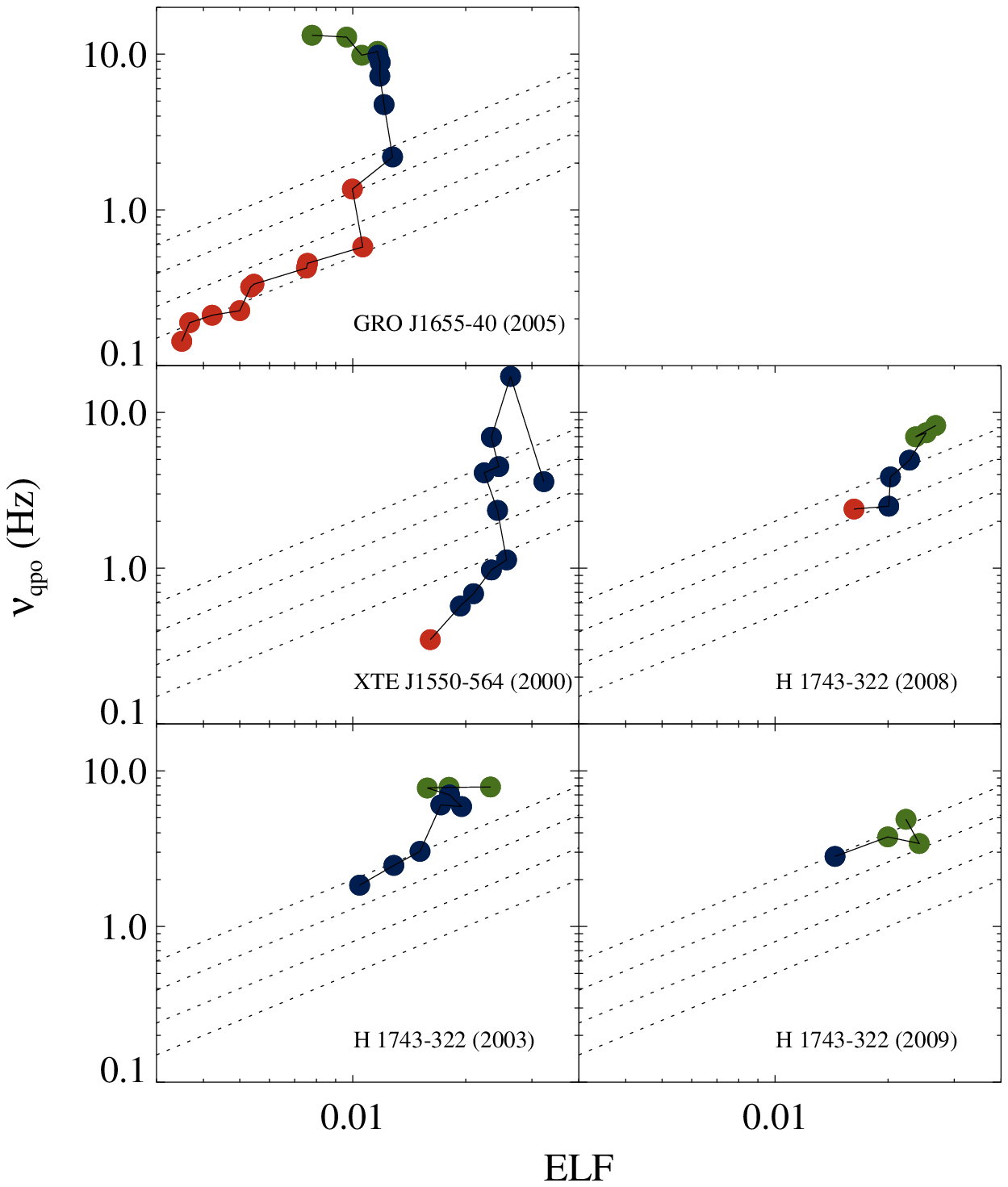}
 \caption{Peak frequency of the QPO vs total ELF for the outliers. Dotted lines show $\nu_{qpo}$ = A $L_{X}^{1.0}$ relation for A = 50, 80, 130, 200.}
 \label{fig:qpoL}
\end{figure}

In Figure~\ref{fig:qpoL}, we plot the peak frequencies of the QPOs against the ELF because the MDAF model predicts a scaling between them in the form of $\nu_{QPO}$ $\propto$ $L_{X}^p$ where $\nu$ is the peak frequency of the QPO, $L_X$ the X-ray luminosity and p = 0.9-1.1 \citep{Meier12}. In this figure, we also show scaling curves with p = 1.0 for various normalizations. The observed relation varies from decay to decay and shows only a partial agreement with the predicted scaling. If we take a look at the sources, GRO~J1655$-$40 and H1743-322 (2003) converge to the predicted scaling at their lower ELF observations, H1743-322 (2009) is in agreement with the scaling, and XTE J1550-564 and H1743-322 (2008) behave far from the predicted scaling.

\section{Discussion}
We performed a thorough analysis of the X-ray variability properties of GBHTs during the decay of their outbursts while the compact jets turn on. When combined with the X-ray spectral information, our results provided strong constraints for the phenomenological understanding of the accretion-ejection process in GBHTs. In the following we discuss our findings.

\subsection{Relation Of X-ray Variability To Radio/X-ray Luminosity Relation}
We showed that the X-ray variability does not reach the same strength for all GBHTs under similar X-ray spectral conditions (ELF and power-law index). Its strength shows a dual behavior connected with the tracks in the radio/X-ray luminosity relation, the standards show higher X-ray variability. This observational connection may be showing us the importance of the X-ray variability on jet emission strength. A recent internal shock model proposed to explain the jet spectral energy distribution supports this idea \citep{Malzac13, Malzac14}. In this model, the jet radio luminosity depends on the amplitude of the Lorentz factor fluctuations of the jet at its base. These fluctuations are related, in some unknown way, to the fluctuations in the accretion flow and therefore to the X-ray variability. While this model does not explain why some sources have higher variability amplitude compared to others, it can explain why sources with higher rms amplitude of variability show higher radio emission and form the standard track in radio/X-ray correlation.

The presence of dual tracks in the radio/X-ray correlation means that either for the given X-ray luminosity there is a difference in radio luminosity or for the given radio luminosity there is a difference in X-ray luminosity. \cite{Meyer-Hofmeister14} favours the latter idea and invokes a cool disk that increases seed photon input for Comptonization thereby resulting in higher X-ray luminosity in outliers. The disk temperature is low, therefore its emission does not show up in the PCA band. \cite{Meyer-Hofmeister14} does not present a prediction for timing properties, and if the outlier track is caused by the cool inner disk, this cool disk must also be affecting short term X-ray timing properties. Another explanation that favours a change in X-ray luminosity could be radiatively efficient accretion in the case of outliers \citep{Coriat11}. On the other hand,  \cite{Casella09} claim that the tracks are caused by a difference in jet magnetic field, such that at a magnetic field greater than a critical value the radio emission is suppressed. Neither \cite{Coriat11} nor \cite{Casella09} offers a prediction on X-ray timing properties, yet for both cases the timing properties may be related to properties of turbulence in the corona which in turn should be related to the magnetic field strength. While we cannot offer any specific explanation for the difference in variability strength for sources in different tracks, it is also clear that models which also explain the difference in timing properties are favored. As a prediction of this work, if the tracks and rms amplitude of variability are directly related, for the cases where the source move from the outlier to the standard track, we should also observe an increase in rms amplitude of variability, and perhaps the differences in rms amplitude of variability will be less at higher X-ray luminosities for which the tracks start to merge. This investigation will be subject of future work.

The only model that attempts to tie variability properties to the tracks in the radio/X-ray correlation is the MDAF model \citep{Meier12}. In this model, the dual tracks can be explained if a thin accretion disk is truncated by the advection dominated accretion flow (ADAF) in the standards and in addition to this, the ADAF itself is truncated at a radius by an MDAF in the outliers. Since the MDAF is a laminar and non-turbulent flow, a high frequency cut-off is expected in the PDS of the outliers as opposed to the ADAF only configuration for the standards. For our dataset, the PDS above 10 Hz usually have large errors. While the drop in the peak frequencies of highest frequency Lorentzians in GRO~J1655$-$40 and XTE~J1550$-$564 (outliers) as opposed to continued presence of a high frequency broad component in GX~339$-$4 (standard) in 2005 and 2007 (see Figure~\ref{fig:timevolfreq}) seems to support the idea of a possible cut-off in outliers at higher frequencies, the quality of the data did not allow us to significantly confirm the predictions. The data in outburst rise with higher statistical quality will be used to test this hypothesis in a future work.

Another prediction of the MDAF theory is that the rotation frequency of the MDAF corresponds to the QPO frequency. However, the QPOs are seen in both the standards (e.g. QPOs in 4U~1543$-$47, \citealt{Kalemci05}) and outliers (QPOs in GRO~J1655$-$40, H1743$-$322, XTE~J1550$-$564). Finally, the predicted scaling between the X-ray luminosity and the QPO frequency by the MDAF model seems to work well only for GRO~J1655$-$40 and H1743-322 (2003 and 2009) at low ELF (see Figure~\ref{fig:qpoL}). This is surprising because the MDAF model is supposed to operate at higher luminosity, between 0.01 and 0.1 of the Eddington Luminosity \citep{Meier12}. We must stress that we calculated the ELF in the limited energy range of 3--200 keV. Before the index transition there may be significant disk contribution below 3 keV, so a proper bolometric calculation (which is not possible with the {\it{RXTE}} data) would have resulted in higher ELF for higher QPO frequencies which could align points through the predicted scale in Figure~\ref{fig:qpoL}. Observations with sufficient low energy coverage are required to verify this prediction. 

\subsection{Connection Between Broadband Noise and QPO}
When the evolution of the QPOs from all the outburst decays are considered together, there is a global trend: the QPO rms amplitude of variability first increases to a maximum value, then the QPO either disappears or its rms amplitude of variability decreases. This is in contrast to the behaviour of the total rms amplitude of variabilities which stay approximately constant after they reach their maximum (see Figure~\ref{fig:totqporms}).
 
The positive correlation between the QPO and the total rms amplitudes after the timing transition suggests that both components strengthen as the emission from the corona becomes stronger with respect to the disk. Such a change can be explained assuming if the disk produces Poisson noise and the corona produces variability \citep{Kalemci04}. Figures~\ref{fig:rmsplr} and \ref{fig:qpormsplr} clearly support this because both the continuum and the QPO rms increases with increasing PLR.

\cite{Yan13} found that the decrease in the QPO rms amplitude of variability coincides with the jet growth for GRS~1915+105 and the authors suggested a possible connection between two events. In our sample, a similar behavior is observed, for all sources the QPO rms amplitude decreased (and for XTE~J1550$-$564 and 4U~1543$-$47 disappeared from the PSD, see Figure~\ref{fig:evolqporms}) within a few days prior to or following CJT. This can be interpreted in two ways. The jet either lowers the QPO rms amplitude by affecting the accretion medium or completely destroy the QPO mechanism resulting in a non-detection of the QPOs. We note that if there is a destruction of the QPO by the jet, we are not able to constrain it strongly because the upper limits on the QPO rms amplitudes are relatively large. Such a possibility deserves to be investigated in the future observing campaigns of the black hole transients with large effective area and high time resolution telescopes such as ASTROSAT \citep{Agrawal06} and LOFT \citep{Feroci12}.

The decrease in the QPO rms amplitude is independent of the total rms amplitude. This lack of dependency may be understood in terms of the mechanisms producing the total and the QPO variabilities. For instance, the broadband variability may be produced via propagating waves in the accretion disk \citep{Lyubarskii97}, and amplified in the corona and the QPO may be produced by precession of the corona as in Lens-Thirring mechanism \citep{Ingram11}. Such a combination of the mechanisms would also explain the increasing part of the rms amplitudes since both mechanisms depend on the establishment of a strong corona, and the non-varying disk would decrease their amplitude when present.

\acknowledgments
T.D. and E.K. acknowledge financial support from T\"{U}B\.{I}TAK through grant 111T222 and FP7 Initial Training Network Black Hole Universe, ITN 215212. J.A.T. acknowledges partial support from NASA Astrophysics Data Analysis Program grant NNX11AF84G. The authors thank all scientist who contributed to the T\"{u}bingen Timing Tools and Craig Markwardt for MPFIT package. The authors thank D. Russell, D. Meier and J. Malzac for fruitful discussions.

{\it Facilities:} \facility{RXTE}, \facility{CTIO:1.5m}.


\begin{thebibliography}{39}
\expandafter\ifx\csname natexlab\endcsname\relax\def\natexlab#1{#1}\fi

\bibitem[{{Agrawal}(2006)}]{Agrawal06}
{Agrawal}, P.~C. 2006, Advances in Space Research, 38, 2989

\bibitem[{{Belloni}(2010)}]{Belloni10}
{Belloni}, T.~M. 2010, in Lecture Notes in Physics, Vol. 794, The Jet Paradigm,
  ed. T.~Belloni (Berlin: Springer), 53

\bibitem[{{Bradt} {et~al.}(1993){Bradt}, {Rothschild}, \& {Swank}}]{Bradt93}
{Bradt}, H.~V., {Rothschild}, R.~E., \& {Swank}, J.~H. 1993, \aaps, 97, 355

\bibitem[{{Buxton} {et~al.}(2012){Buxton}, {Bailyn}, {Capelo}, {Chatterjee},
  {Din{\c c}er}, {Kalemci}, \& {Tomsick}}]{Buxton12}
{Buxton}, M.~M., {Bailyn}, C.~D., {Capelo}, H.~L., {et~al.} 2012, \aj, 143, 130

\bibitem[{{Casella} \& {Pe'er}(2009)}]{Casella09}
{Casella}, P., \& {Pe'er}, A. 2009, \apjl, 703, L63

\bibitem[{{Chun} {et~al.}(2013){Chun}, {Din{\c c}er}, {Kalemci}, {G{\"u}ver},
  {Tomsick}, {Buxton}, {Brocksopp}, {Corbel}, \& {Cabrera-Lavers}}]{Chun13}
{Chun}, Y.~Y., {Din{\c c}er}, T., {Kalemci}, E., {et~al.} 2013, \apj, 770, 10

\bibitem[{{Corbel} {et~al.}(2013){Corbel}, {Coriat}, {Brocksopp}, {Tzioumis},
  {Fender}, {Tomsick}, {Buxton}, \& {Bailyn}}]{Corbel13}
{Corbel}, S., {Coriat}, M., {Brocksopp}, C., {et~al.} 2013, \mnras, 428, 2500

\bibitem[{{Corbel} {et~al.}(2000){Corbel}, {Fender}, {Tzioumis}, {Nowak},
  {McIntyre}, {Durouchoux}, \& {Sood}}]{Corbel00}
{Corbel}, S., {Fender}, R.~P., {Tzioumis}, A.~K., {et~al.} 2000, \aap, 359, 251

\bibitem[{{Coriat} {et~al.}(2011){Coriat}, {Corbel}, {Prat}, {Miller-Jones},
  {Cseh}, {Tzioumis}, {Brocksopp}, {Rodriguez}, {Fender}, \&
  {Sivakoff}}]{Coriat11}
{Coriat}, M., {Corbel}, S., {Prat}, L., {et~al.} 2011, \mnras, 414, 677

\bibitem[{{Din{\c c}er} {et~al.}(2012){Din{\c c}er}, {Kalemci}, {Buxton},
  {Bailyn}, {Tomsick}, \& {Corbel}}]{Dincer12}
{Din{\c c}er}, T., {Kalemci}, E., {Buxton}, M.~M., {et~al.} 2012, \apj, 753, 55

\bibitem[{{Done} {et~al.}(2007){Done}, {Gierli{\'n}ski}, \& {Kubota}}]{Done07}
{Done}, C., {Gierli{\'n}ski}, M., \& {Kubota}, A. 2007, \aapr, 15, 1

\bibitem[{{Fender} {et~al.}(2001){Fender}, {Hjellming}, {Tilanus}, {Pooley},
  {Deane}, {Ogley}, \& {Spencer}}]{Fender01}
{Fender}, R.~P., {Hjellming}, R.~M., {Tilanus}, R.~P.~J., {et~al.} 2001,
  \mnras, 322, L23

\bibitem[{{Fender} {et~al.}(2009){Fender}, {Homan}, \& {Belloni}}]{Fender09}
{Fender}, R.~P., {Homan}, J., \& {Belloni}, T.~M. 2009, \mnras, 396, 1370

\bibitem[{{Feroci} {et~al.}(2012){Feroci}, {Stella}, {van der Klis},
  {Courvoisier}, {Hernanz}, {Hudec}, {Santangelo}, {Walton}, {Zdziarski},
  {Barret}, {Belloni}, {Braga}, {Brandt}, {Budtz-J{\o}rgensen}, {Campana}, {den
  Herder}, {Huovelin}, {Israel}, {Pohl}, {Ray}, {Vacchi}, {Zane}, {Argan},
  {Attin{\`a}}, {Bertuccio}, {Bozzo}, {Campana}, {Chakrabarty}, {Costa}, {De
  Rosa}, {Del Monte}, {Di Cosimo}, {Donnarumma}, {Evangelista}, {Haas},
  {Jonker}, {Korpela}, {Labanti}, {Malcovati}, {Mignani}, {Muleri},
  {Rapisarda}, {Rashevsky}, {Rea}, {Rubini}, {Tenzer}, {Wilson-Hodge},
  {Winter}, {Wood}, {Zampa}, {Zampa}, {Abramowicz}, {Alpar}, {Altamirano},
  {Alvarez}, {Amati}, {Amoros}, {Antonelli}, {Artigue}, {Azzarello},
  {Bachetti}, {Baldazzi}, {Barbera}, {Barbieri}, {Basa}, {Baykal}, {Belmont},
  {Boirin}, {Bonvicini}, {Burderi}, {Bursa}, {Cabanac}, {Cackett}, {Caliandro},
  {Casella}, {Chaty}, {Chenevez}, {Coe}, {Collura}, {Corongiu}, {Covino},
  {Cusumano}, {D'Amico}, {Dall'Osso}, {De Martino}, {De Paris}, {Di Persio},
  {Di Salvo}, {Done}, {Dov{\v c}iak}, {Drago}, {Ertan}, {Fabiani}, {Falanga},
  {Fender}, {Ferrando}, {Della Monica Ferreira}, {Fraser}, {Frontera},
  {Fuschino}, {Galvez}, {Gandhi}, {Giommi}, {Godet}, {G{\"o}{\v g}{\"u}{\c s}},
  {Goldwurm}, {G{\"o}tz}, {Grassi}, {Guttridge}, {Hakala}, {Henri}, {Hermsen},
  {Horak}, {Hornstrup}, {in't Zand}, {Isern}, {Kalemci}, {Kanbach}, {Karas},
  {Kataria}, {Kennedy}, {Klochkov}, {Klu{\'z}niak}, {Kokkotas}, {Kreykenbohm},
  {Krolik}, {Kuiper}, {Kuvvetli}, {Kylafis}, {Lattimer}, {Lazzarotto}, {Leahy},
  {Lebrun}, {Lin}, {Lund}, {Maccarone}, {Malzac}, {Marisaldi}, {Martindale},
  {Mastropietro}, {McClintock}, {McHardy}, {Mendez}, {Mereghetti}, {Miller},
  {Mineo}, {Morelli}, {Morsink}, {Motch}, {Motta}, {Mu{\~n}oz-Darias},
  {Naletto}, {Neustroev}, {Nevalainen}, {Olive}, {Orio}, {Orlandini},
  {Orleanski}, {Ozel}, {Pacciani}, {Paltani}, {Papadakis}, {Papitto},
  {Patruno}, {Pellizzoni}, {Petr{\'a}{\v c}ek}, {Petri}, {Petrucci}, {Phlips},
  {Picolli}, {Possenti}, {Psaltis}, {Rambaud}, {Reig}, {Remillard},
  {Rodriguez}, {Romano}, {Romanova}, {Schanz}, {Schmid}, {Segreto}, {Shearer},
  {Smith}, {Smith}, {Soffitta}, {Stergioulas}, {Stolarski}, {Stuchlik},
  {Tiengo}, {Torres}, {T{\"o}r{\"o}k}, {Turolla}, {Uttley}, {Vaughan},
  {Vercellone}, {Waters}, {Watts}, {Wawrzaszek}, {Webb}, {Wilms}, {Zampieri},
  {Zezas}, \& {Ziolkowski}}]{Feroci12}
{Feroci}, M., {Stella}, L., {van der Klis}, M., {et~al.} 2012, Experimental
  Astronomy, 34, 415

\bibitem[{{Gallo} {et~al.}(2012){Gallo}, {Miller}, \& {Fender}}]{Gallo12}
{Gallo}, E., {Miller}, B.~P., \& {Fender}, R. 2012, \mnras, 423, 590

\bibitem[{{Homan} {et~al.}(2005){Homan}, {Buxton}, {Markoff}, {Bailyn},
  {Nespoli}, \& {Belloni}}]{Homan05}
{Homan}, J., {Buxton}, M., {Markoff}, S., {et~al.} 2005, \apj, 624, 295

\bibitem[{{Ingram} \& {Done}(2011)}]{Ingram11}
{Ingram}, A., \& {Done}, C. 2011, \mnras, 415, 2323

\bibitem[{{Kalemci} {et~al.}(2013){Kalemci}, {Din{\c c}er}, {Tomsick},
  {Buxton}, {Bailyn}, \& {Chun}}]{Kalemci13}
{Kalemci}, E., {Din{\c c}er}, T., {Tomsick}, J.~A., {et~al.} 2013, \apj, 779,
  95

\bibitem[{{Kalemci} {et~al.}(2005){Kalemci}, {Tomsick}, {Buxton}, {Rothschild},
  {Pottschmidt}, {Corbel}, {Brocksopp}, \& {Kaaret}}]{Kalemci05}
{Kalemci}, E., {Tomsick}, J.~A., {Buxton}, M.~M., {et~al.} 2005, \apj, 622, 508

\bibitem[{{Kalemci} {et~al.}(2003){Kalemci}, {Tomsick}, {Rothschild},
  {Pottschmidt}, {Corbel}, {Wijnands}, {Miller}, \& {Kaaret}}]{Kalemci03}
{Kalemci}, E., {Tomsick}, J.~A., {Rothschild}, R.~E., {et~al.} 2003, \apj, 586,
  419

\bibitem[{{Kalemci} {et~al.}(2004){Kalemci}, {Tomsick}, {Rothschild},
  {Pottschmidt}, \& {Kaaret}}]{Kalemci04}
{Kalemci}, E., {Tomsick}, J.~A., {Rothschild}, R.~E., {Pottschmidt}, K., \&
  {Kaaret}, P. 2004, \apj, 603, 231

\bibitem[{{Klein-Wolt} \& {van der Klis}(2008)}]{Klein-Wolt08}
{Klein-Wolt}, M., \& {van der Klis}, M. 2008, \apj, 675, 1407

\bibitem[{{Lyubarskii}(1997)}]{Lyubarskii97}
{Lyubarskii}, Y.~E. 1997, \mnras, 292, 679

\bibitem[{{Malzac}(2013)}]{Malzac13}
{Malzac}, J. 2013, \mnras, 429, L20

\bibitem[{{Malzac}(2014)}]{Malzac14}
---. 2014, ArXiv e-prints

\bibitem[{{Meier}(2012)}]{Meier12}
{Meier}, D.~L. 2012, {Black Hole Astrophysics: The Engine Paradigm}

\bibitem[{{Meyer-Hofmeister} \& {Meyer}(2014)}]{Meyer-Hofmeister14}
{Meyer-Hofmeister}, E., \& {Meyer}, F. 2014, ArXiv e-prints

\bibitem[{{Miller-Jones} {et~al.}(2012){Miller-Jones}, {Sivakoff},
  {Altamirano}, {Coriat}, {Corbel}, {Dhawan}, {Krimm}, {Remillard}, {Rupen},
  {Russell}, {Fender}, {Heinz}, {K{\"o}rding}, {Maitra}, {Markoff}, {Migliari},
  {Sarazin}, \& {Tudose}}]{Miller12}
{Miller-Jones}, J.~C.~A., {Sivakoff}, G.~R., {Altamirano}, D., {et~al.} 2012,
  \mnras, 421, 468

\bibitem[{{Miyamoto} \& {Kitamoto}(1989)}]{Miyamoto89}
{Miyamoto}, S., \& {Kitamoto}, S. 1989, \nat, 342, 773

\bibitem[{{Ratti} {et~al.}(2012){Ratti}, {Jonker}, {Miller-Jones}, {Torres},
  {Homan}, {Markoff}, {Tomsick}, {Kaaret}, {Wijnands}, {Gallo}, {{\"O}zel},
  {Steeghs}, \& {Fender}}]{Ratti12}
{Ratti}, E.~M., {Jonker}, P.~G., {Miller-Jones}, J.~C.~A., {et~al.} 2012,
  \mnras, 423, 2656

\bibitem[{{Remillard} \& {McClintock}(2006)}]{Remillard06}
{Remillard}, R.~A., \& {McClintock}, J.~E. 2006, \araa, 44, 49

\bibitem[{{Russell} {et~al.}(2010){Russell}, {Maitra}, {Dunn}, \&
  {Markoff}}]{Russell10}
{Russell}, D.~M., {Maitra}, D., {Dunn}, R.~J.~H., \& {Markoff}, S. 2010,
  \mnras, 405, 1759

\bibitem[{{Russell} {et~al.}(2011){Russell}, {Miller-Jones}, {Maccarone},
  {Yang}, {Fender}, \& {Lewis}}]{Russell_q11}
{Russell}, D.~M., {Miller-Jones}, J.~C.~A., {Maccarone}, T.~J., {et~al.} 2011,
  \apjl, 739, L19

\bibitem[{{Shakura} \& {Sunyaev}(1973)}]{Shakura73}
{Shakura}, N.~I., \& {Sunyaev}, R.~A. 1973, \aap, 24, 337

\bibitem[{{Soleri} \& {Fender}(2011)}]{Soleri11}
{Soleri}, P., \& {Fender}, R. 2011, \mnras, 413, 2269

\bibitem[{{Tanaka} \& {Shibazaki}(1996)}]{Tanaka96}
{Tanaka}, Y., \& {Shibazaki}, N. 1996, \araa, 34, 607

\bibitem[{{Vaughan} {et~al.}(2011){Vaughan}, {Uttley}, {Pounds}, {Nandra}, \&
  {Strohmayer}}]{Vaughan11}
{Vaughan}, S., {Uttley}, P., {Pounds}, K.~A., {Nandra}, K., \& {Strohmayer},
  T.~E. 2011, \mnras, 413, 2489

\bibitem[{{Yan} {et~al.}(2013){Yan}, {Ding}, {Wang}, {Qu}, \& {Song}}]{Yan13}
{Yan}, S.-P., {Ding}, G.-Q., {Wang}, N., {Qu}, J.-L., \& {Song}, L.-M. 2013,
  \mnras

\bibitem[{{Zhang} {et~al.}(1995){Zhang}, {Jahoda}, {Swank}, {Morgan}, \&
  {Giles}}]{Zhang95}
{Zhang}, W., {Jahoda}, K., {Swank}, J.~H., {Morgan}, E.~H., \& {Giles}, A.~B.
  1995, \apj, 449, 930

\end{thebibliography}

\end{document}